\documentclass{article}
\usepackage{waspaa19,amssymb,amsmath,graphicx,times,url}
\usepackage{color}
\usepackage{cite}

\title{GENERATIVE SPEECH ENHANCEMENT BASED ON CLONED NETWORKS}

\name{Michael Chinen,$^{1}$ 
      W. Bastiaan Kleijn,$^{1,2}$ 
      Felicia S. C. Lim,$^{1}$ 
      Jan Skoglund$^{1}$}
\address{$^1$ Google LLC, USA\\
         $^2$ School of Engineering and Computer Science, Victoria University of Wellington, New Zealand
}

\begin{document}

\ninept
\maketitle

\begin{sloppy}

\begin{abstract}
We propose to implement speech enhancement by the regeneration of clean speech from a `salient' representation extracted from the noisy signal. The network that extracts salient features is trained using a set of weight-sharing clones of the extractor network. The clones receive mel-frequency spectra of different noisy versions of the same speech signal as input. By encouraging the outputs of the clones to be similar for these different input signals, we train a feature extractor network that is robust to noise. At inference, the salient features form the input to a WaveNet network that generates a natural and clean speech signal with the same attributes as the ground-truth clean signal. As the signal becomes noisier, our system produces natural sounding errors that stay on the speech manifold, in place of traditional artifacts found in other systems.  Our experiments confirm that our generative enhancement system provides state-of-the-art enhancement performance within the generative class of enhancers according to a MUSHRA-like test.  The clones based system matches or outperforms the other systems at each input signal-to-noise (SNR) range with statistical significance.

\end{abstract}


\section{Introduction}
\label{sec:intro}

The performance of speech enhancement has 
leaped significantly with the introduction of deep neural networks (DNNs) in recent years, e.g., \cite{yoshioka2015ntt,erdogan2015phase,xu2015regression,kumar2016speech,park2017fully,valentini2016investigating,wilson2018exploring,wisdom2019differentiable,zhao2018convolutional,luo2018tasnet,rethage2018wavenet,pascual2017segan}. This advance can be attributed to DNNs being unencumbered by explicit or implicit constraints on relevant data probability distributions, replacing these distributions with empirical observations, and by the ability of DNNs to capture complex relationships that cannot be expressed analytically. The goal of this paper is to explore a new class of enhancement, where the clean signal is generated from features that are extracted from contaminated signals using an extraction network. 

In most work on enhancement, e.g., \cite{loizou2007speech,benesty2006speech,gannot2001signal,mohammadiha2013supervised}, the observed signal is considered to be an additive mixture of a clean speech signal ${\bf x}$ and additional noise sources:
\begin{equation}
    {\bf y} = {\bf x} + \sum_{j=1}^J {\bf n}_j .
    \label{eq:mix}
\end{equation}
It follows from \eqref{eq:mix} that a natural objective for enhancement is to find a good approximation of the clean speech waveform ${\bf x}$ based on the noisy observations $\bf{y}$ and available prior knowledge.
Optimizing a network to minimize a measure of error between a clean signal estimate $\hat{\bf x}$ and the ground-truth ${\bf x}$ is a common approach that has yielded state of the art results based on DNN approaches operating in the  time-frequency domain \cite{erdogan2015phase,xu2015regression,kumar2016speech,park2017fully,valentini2016investigating,wilson2018exploring,wisdom2019differentiable} or time-only domain \cite{luo2018tasnet}. Typically, a feed-forward or Recurrent Neural Network (RNN) is used.

Instead of feed-forward networks, generative networks have also been used. Recent generative systems such as WaveNet \cite{van2016wavenet} and Generative Adversarial Networks (GAN) \cite{goodfellow2014generative} have shown excellent results for speech synthesis and image generation. It is natural to exploit these advanced signal models in speech enhancement. Examples of this approach are  \textit{a WaveNet for speech denoising} \cite{rethage2018wavenet}, SEGAN \cite{pascual2017segan}, and related GAN-based approaches \cite{michelsanti2017conditional, donahue2018exploring}.  These applications borrow the architecture from the generative systems they are named after. However, they are at least in part optimized to reconstruct the ground-truth waveform, thus restricting the generative aspect of the system.

Typical speech enhancement systems tend to degrade into characteristic artifacts that sound unnatural as the noise level increases.  Wiener filters do go quiet; it is their uneven way of doing this that enhances and also causes distortion.  Generative systems have the property that they are restricted to producing natural speech, which may be interpreted as being restricted to a ``speech manifold'', and provides an interesting solution to this problem. For example, WaveNets trained to produce speech should have difficulty reproducing non-speech sounds and will tend to produce phoneme or inebriation-like errors in place of artifacts as noise levels increase. 

Generative models can use a stochastic process to generate complex details that are perceptually irrelevant instead of trying to reproduce the input signal exactly.  There may be many plausible solutions that differ considerably from the ground truth but are equivalent to it.  For example, the phase in noisy signals, prosody in text-to-speech, or the texture of sand in imaging may have many acceptable solutions that a generative system can model and exploit with a more compact representation.  Traditional enhancement metrics for evaluation, including signal-to-noise ratio (SNR) and source-to-distortion/interference/artifacts ratios (SDR/SIR/SAR) \cite{vincent2006performance} use a reference, hence discouraging many plausible solutions, and will therefore not be effective metrics for generative models.  Perceptual speech models such as PESQ \cite{rix2001perceptual} and ViSQOL \cite{hines2012visqol} are more flexible in this regard, but still rely too heavily on the degraded signal matching a provided reference.  As a result, human evaluation is the standard metric we propose for true generative enhancement \cite{kleijn2018wavenet}.

Our contribution is to use cloned networks trained on many noisy copies of a given utterance to extract a representation of speech, which is similar across all clones and contains only the information needed to reconstruct the shared speech utterance, discarding information about the noise.  These features are used as conditioning to a generative network.  To encourage the network to find useful information we use a decoder network with a loss targeting the clean, mel-frequency spectra.  This deterministic decoder network mirrors the extractor, and is independent of the WaveNet network used for inference. The method can function without a decoder, but the decoder boosts performance significantly \cite{kleijn2019clones}.

The outline of this paper is as follows. In section \ref{sec:clone_based_enhancement} we provide a detailed motivation for \textit{generative} enhancement and describe the clone-based training procedure. Section \ref{sec:experiments} describes the experiments and their outcomes. Finally section \ref{sec:conclusion} provides conclusions drawn from our work.

\section{CLONE-BASED ENHANCEMENT}
\label{sec:clone_based_enhancement}
In this section we first provide our motivation for generative enhancement. We then outline the architecture of the extraction network and discuss in more detail the clone-based training setup.

\subsection{Generative Enhancement}
\label{ssec:generative_systems}
The enhancement of single-channel speech has long been aimed at finding an approximation of the clean signal waveform, typically measured with an objective function that can be interpreted as a Bayes risk. Early approaches used a Wiener filter to this purpose, which meant that noise reduction was associated with signal distortion. More recent methods are based on models of the speech and/or noise signals. The prior knowledge implicit in the models facilitates a better trade-off between noise reduction and signal distortion. It can be argued that the waveform distortion resulting from classical methods is a direct result of attempting to extract the clean signal waveform based on a Bayes risk: waveform components that cannot be rendered accurately are reduced in amplitude. 

Recently, machine-learning has enabled the rendering of natural sounding speech based only on a parameter sequence with an information rate that is near the rate of the linguistic message conveyed. This suggests a new enhancement paradigm. If a suitable parameter sequence can be extracted from the noisy signal, then an equivalent natural-sounding speech signal can be constructed. This synthesized signal is, in general, not a reconstruction of the original clean signal waveform; it has a high distortion according to conventional measures. A characteristic of the generative approach is that, with current methods, the waveform of the rendered signal is not unique, relying on the drawing of samples from a statistical distribution that can be interpreted as an information generating step.

Perhaps surprisingly, enhancement methods based on generative networks generally are either based solely on attempting to reconstruct the signal waveform or rely on this objective at least in part. That is, these methods exploit the sophistication of the generative network model to find a better approximation of the clean signal waveform. For example, \cite{qian2017speech} approximates the clean signal waveform using a Bayesian formalism that incorporates the structure of WaveNet.  \cite{rethage2018wavenet} uses a WaveNet structure to create a deterministic mapping from the noisy waveform to the clean waveform approximation. The SEGAN approach \cite{pascual2017segan} combines a generative adversarial approach with an L1 criterion that encourages reconstruction of the signal waveform. The SEGAN reconstruction is further facilitated by skip connections that allow information about the waveform to be passed directly to the decoder.
 
 In this paper we explore generative enhancement without attempting to approximate the clean signal waveform. Our aim is (i) to show that this approach is a practical enhancement method and (ii) to explore the behavior of a generative system. We would expect that the degradation with decreasing SNR is different from methods that attempt to approximate the clean waveform. When the clean waveform is approximated, the reconstruction will become increasingly unnatural and decreases in energy with decreasing SNR. In contrast, we expect that the generative system will generate increasingly unintelligible speech with decreasing SNR, with errors in phones that are obscured by the noise to appear first. We will confirm this behavior in section \ref{sec:experiments}.
 
 \subsection{Network Architecture}
\label{ssec:network_architecture}
The basic architecture of our network, as used for inference, consists of a feature extraction network, which extracts a time sequence of salient-feature vectors. It can be used as the conditioning for a conventional WaveNet generative network \cite{van2016wavenet}. The salient features are selected in a training process, but based on qualitative designer input.

\begin{figure}[t]
  \centering
  \centerline{\includegraphics[width=\columnwidth]{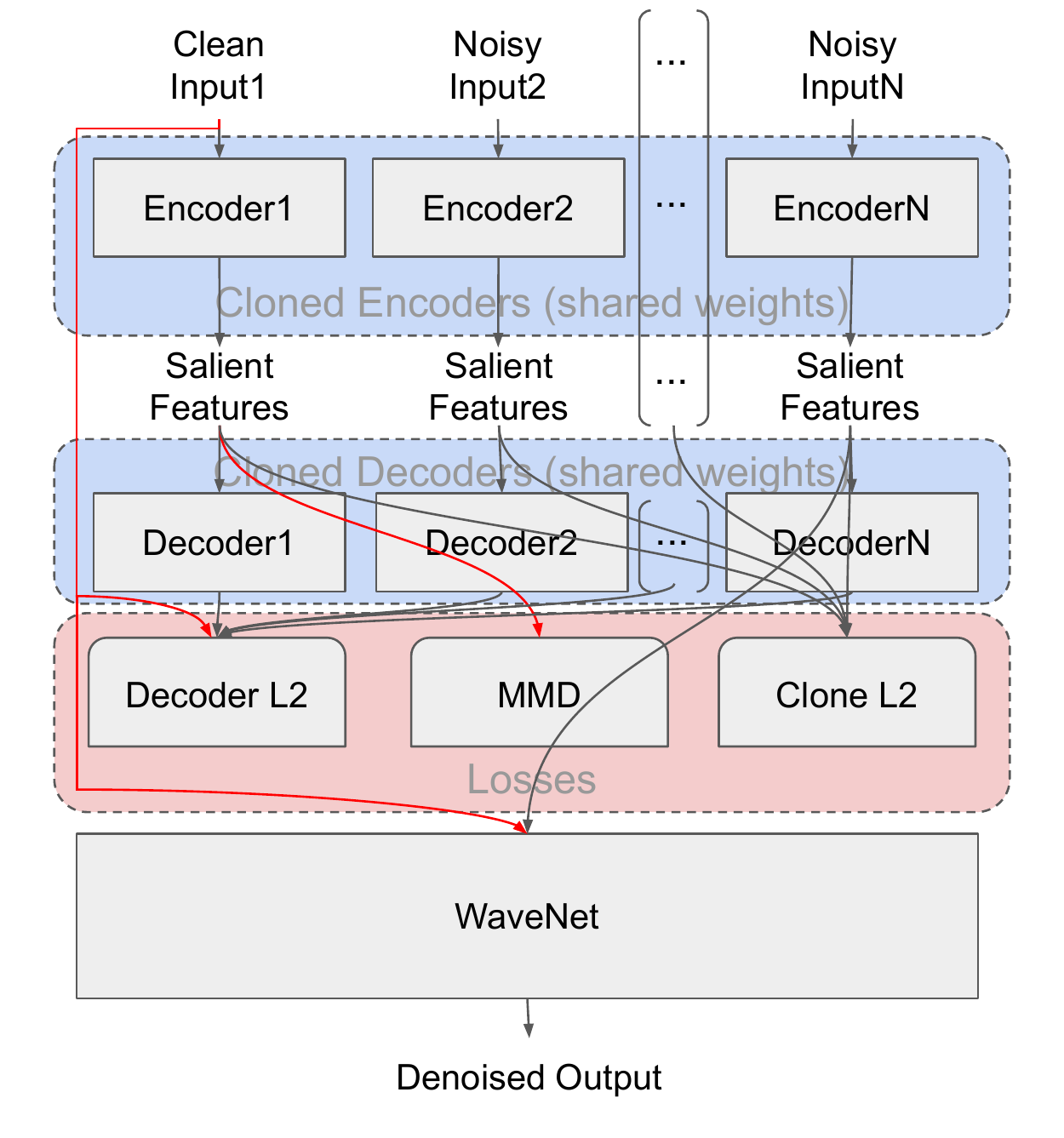}}
  \caption{The basic architecture of our system.  Red lines show where only the first clone is used.}
  \label{f:trainingSetup}
\end{figure}

Before the training process is started, the system designer selects a set of \textit{equivalent} signals that share only the salient features. For example, the designer might add different noise to the same speech utterance. Alternatively,  the same utterance could be passed through different all-pass filters and/or be subjected to random delays.

The salient features are found by encouraging \textit{clones} of the extraction network to extract the same feature vectors from different equivalent signals. This leads to the training setup shown in Fig. \ref{f:trainingSetup}. To ensure that the individual features are distinct, we encourage the feature vector components to be independent and have unit variance. We additionally encourage the components to be sparse. To improve performance further, we add a decoder that attempts to reconstruct a target signal derived from a clean equivalent signal.

Our global loss function used for training has three terms. They relate to equivalence, the triple independence, unit-variance and sparsity, and target signal reconstruction, respectively:
\begin{equation}
D_{\mathrm{global}} = D_E + \lambda_{\mathrm{MMD}} D_{\mathrm{MMD}} + \lambda_D D_D,
\label{q:globalLoss}
\end{equation}
where $\lambda_{\mathrm{MMD}}$ and $\lambda_D$ are weightings. 
Let the encoder network $g_\psi: \mathbb{R}^N \rightarrow \mathbb{R}^L$ be the deep network based mapping from a block of data of dimensionality $N$ to the feature vector $z \in \mathbb{R}^L$ with parameters (weights) $\psi$ and $f_\phi: \mathbb{R}^L \rightarrow \mathbb{R}^N$ is the decoder network with learned parameters $\phi$ . Then our objective is to minimize $D_{\mathrm{global}}$ by selecting a suitable set $\psi$ and $\phi$. It is natural to define the loss $D_{\mathrm{global}}$ as an expectation over the data distribution. For a practical application, we instead write $D_{\mathrm{global}}$ as an average over observed batch of $m$ data. We then use stochastic gradient descent over the batches to optimize over the empirical distribution associated with the database at hand.

 The term that encourages the extraction of the same feature set from equivalent input signals is then:
\begin{equation}
    D_E = \sum_{i=1}^m \sum_{q=2}^{q=Q} \| z_i^{(1)} - z_i^{(q)} \|^2,
    \label{q:equivalent}
\end{equation}
where $z_i^{(q)}\in \mathbb{R}^L$ is the feature set of clone $q$, $m$ is the data batch cardinality, $\|\cdot \|$ is the 2-norm, and where we elected to use the first clone as reference to reduce computational effort. An alternative is to compare all clones with all other clones.

The second term of the  global loss function is to encourage independence of the features, unit variance, and sparsity. We implement this objective by encouraging the features vectors to have a suitable distribution: a multivariate Laplacian with independent components and unit variance. For the second term we selected the maximum mean discrepancy (MMD) measure \cite{borgwardt2006integrating}. MMD is a measure of the difference between two distributions. For a batch of $m$ training data blocks\footnote{A training data entry corresponds to a data block (frame). For example 40 ms of speech.} with $y_i$ drawn from the desired Laplacian distribution and $x_i$ the observed data, the square of the MMD measure can be approximated as
\begin{multline}
   D_{\mathrm{MMD}}= \frac{1}{m(m-1)} \sum_{i\neq j}^m \left( k(z_i,z_j) 
    +  \ k(y_i,y_j) \right) \\
    - \frac{2}{m^2}\sum_{i, j}^{m,m} k(z_i,y_j),
    \label{q:mmd}
\end{multline}
where $k(\cdot,\cdot)$ is a kernel with suitably selected scale and shape. We use the multiquadratic kernel \cite{tolstikhin2017wasserstein} and generally operate the MMD measure on only one clone. Note that $m$ must be sufficiently large to make the MMD perform correctly, even for a slow learning rate.

The third term of the global loss function relates to the reconstruction of a target signal derived from a clean equivalent signal. To remain consistent with our notion of generative enhancement, we must, in general, not attempt to approximate the signal waveform directly. However, it is consistent to train our feature extraction clones in tandem with the WaveNet training structure.  In practice, it is more straightforward to use a target signal that characterizes the short-term spectral characteristic of the signal as that requires lower computational effort. We use a L2 norm for this purpose; the decoder loss function is:
\begin{equation}
    D_D = \sum_{i=1}^m \sum_{q=1}^Q\| f_\phi(z_i^{(q)}) - v_i\|^2,
    \label{q:decoderloss}
\end{equation}
where $v_i\in\mathbb{R}^N$ is a suitable signal representation for a block of clean data, $f_\phi$ is the decoder network, and where the summation is over the $m$ batch elements. 

The deterministic decoder loss is only used to help extract the salient features in the clones network, and should not limit the generative capacity of WaveNet.   Additionally, equivalent clean signals in the noisy mixture may not be identical for each clone (e.g., they may differ by time-shifts, or even be from different recordings of the same utterance spoken in an equivalent manner). The decoder loss target has no single or specific ground truth, and this term can be viewed as a metric for determining if the information to reconstruct a particular solution exists in the salient features from any equivalent clean signal.
   
The SNR of the speech signals that are used for training is of critical importance for the performance of the feature extraction. If the signals provided to the clones are too noisy, then the objective \eqref{q:equivalent} will encourage the removal of speech attributes that are perceived by humans and relevant to comprehension and/or quality. On the other hand, if the signals provided to the clones are insufficiently noisy, then the extraction procedure will result in features that are sensitive to noise.

\section{EXPERIMENTS}
\label{sec:experiments}
In this section we first describe the experimental setup of the listening test, then the configuration of the networks, and finally the experimental results.
\subsection{Experimental setup}
\label{ssec:experimental_setup}
\subsubsection{Network Parameters}

The scale of the kernel in \eqref{q:mmd} was selected as 1.0. The weightings $\lambda_{\mathrm{MMD}}$ and $\lambda_D$ in \eqref{q:globalLoss} were set to 1.0 and 18.0, respectively.

We trained two sizes of models that encoded the features.  The \textit{small} model used two layers of Long Short-Term Memory \cite{hochreiter1997long} (LSTM) cells followed by one fully connected layer, each with 800 nodes, and the \textit{large} model used three layers of LSTMs and two layers of fully connected nodes, also with 800 nodes per layer. A final fully-connected layer in both models used linear activation to avoid saturation and to reduce the dimensionality to the desired size. A mirror of this was used for the decoder.

Each training step processed a sequence from $16$~kHz input of six 40-ms dual-window mel-frequency frames that overlapped by 50 percent.  Each of the dual-window frames consist of 80 mel-frequency bins from one window of 40 ms and two windows of $20$~ms (located at 5-$25$~ms, and 15-$45$~ms of the 40 ms window) for a total of 240 mel-frequency bins per frame.  The dual-window approach was used because it was found to produce less consonant phoneme confusion.

 The encoder network used 32 clones and a batch size of 64.  Each clone was constructed of a clean utterance, plus a different noise (from within the dataset).  The encoder network outputs 12 salient features per frame. The salient features were then inferred on full utterances from the training set to create the conditioning training data for WaveNet, using the clean speech for teacher forcing and negative log-likelihood loss.

\subsubsection{Dataset}
We chose to use the VoiceBank-DEMAND speech enhancement dataset provided by Valentini et al. \cite{valentini2016investigating}, because it is public and because other systems have been evaluated on it, providing a useful benchmark.  Because of the relatively small size of the the data, to prevent overfitting we trained our models and picked the checkpoint with the smallest loss. We trained for approximately 200,000 steps for the clones model and 129,000 steps for our WaveNet models.

\subsubsection{Listening Test}
We conducted MUSHRA-like listening tests with 100 listeners per example to evaluate the performance of our system. The reference was the original clean utterance, and the noisy signals were used as anchors. 

We conducted separate tests to evaluate our models against SEGAN and \textit{a WaveNet for Denoising} (labeled as SEWaveNet). We applied our system to the noisy signals from the public examples from both systems to allow a fair comparison.  Both SEWaveNet and SEGAN provided 20 utterances (which did not overlap enough to conduct a single test).  The utterances for each system were provided in different input-SNR ranges.  As additional perspective, both SEGAN and SEWaveNet were shown to outperform a Wiener filter in their respective studies.

\subsection{Results}
\label{ssec:results}

\begin{figure}[t]
  \centering
  \centerline{\includegraphics[width=\columnwidth]{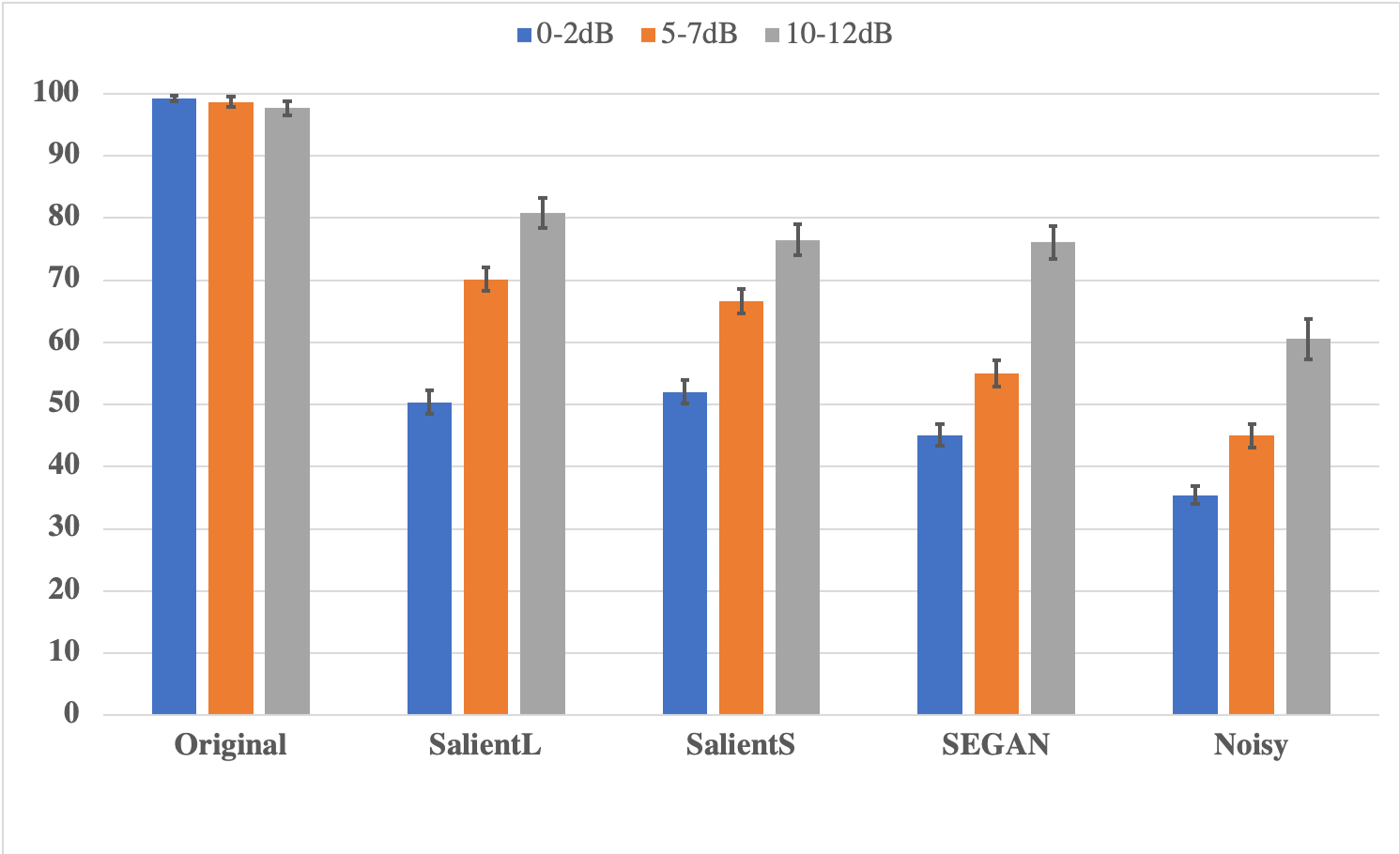}}
  \caption{Results comparing clone-based enhancement to SEGAN \cite{pascual2017segan}. Different colors are used for different input-SNR ranges.}
  \label{fig:segan-results}
\end{figure}
\begin{figure}[t]
  \centering
  \centerline{\includegraphics[width=\columnwidth]{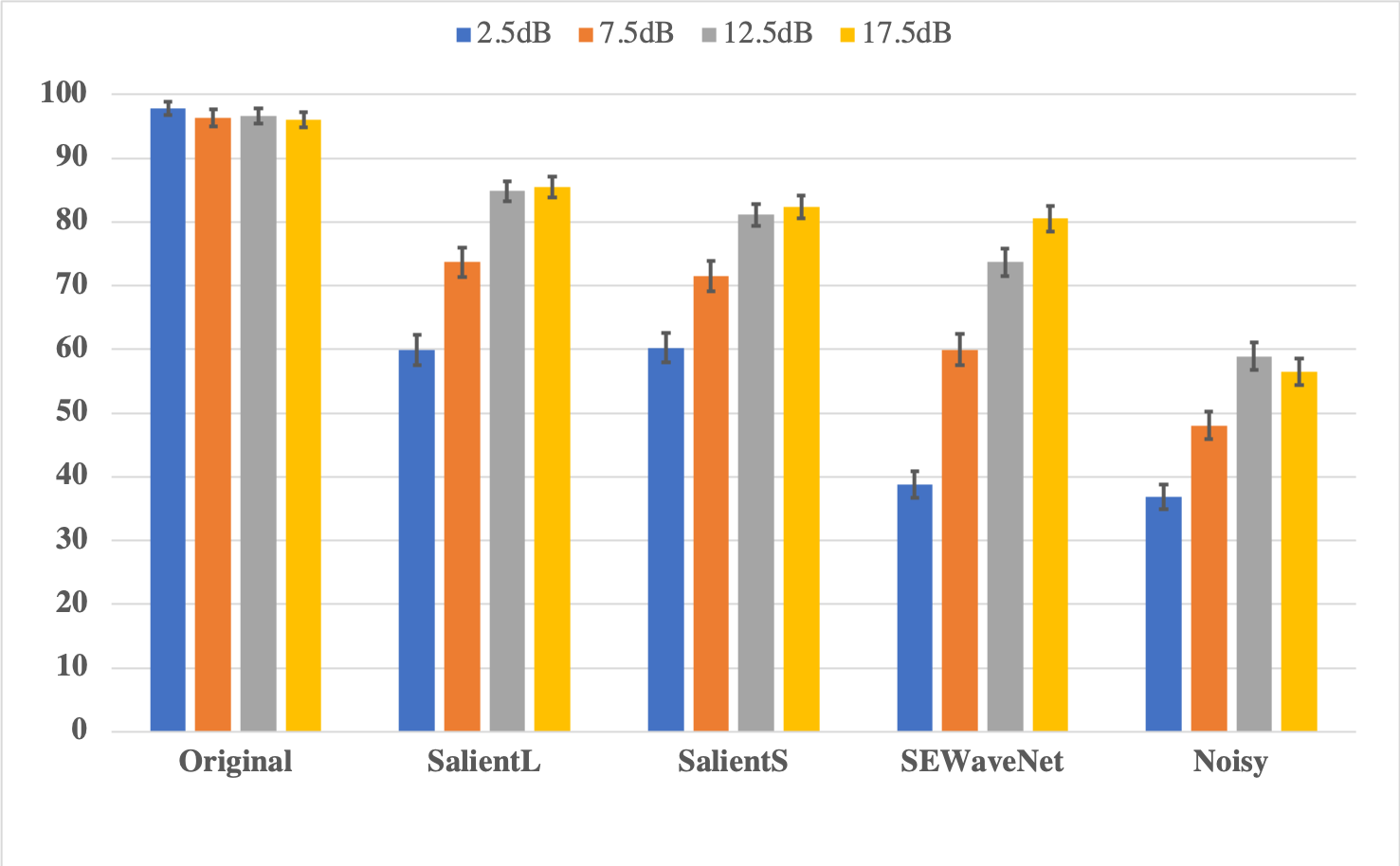}}
  \caption{Results comparing clone-based enhancement to SEWaveNet \cite{rethage2018wavenet}.}
  \label{fig:sewavenet-results}
\end{figure}

\begin{figure}[!htb]
  \centering
  \centerline{\includegraphics[width=\columnwidth]{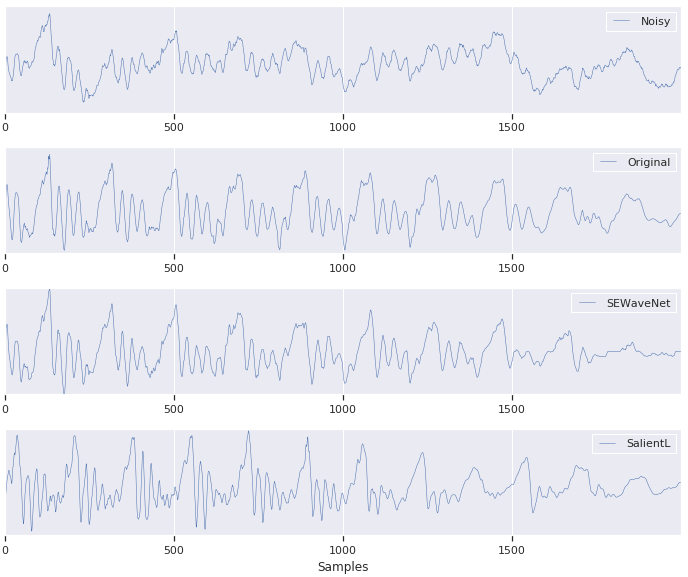}}
  \caption{Waveforms comparing the generated speech by our system to the reference, noisy and SEWavenet \cite{rethage2018wavenet}.  The referential L1 loss from SEWaveNet requires that it match the phase of the reference, unlike our system, which produces significantly different waveforms.}
  \label{fig:waveform-comparison}
\end{figure}

The results for the experiment against SEGAN and SEWaveNet can be seen in Fig.~\ref{fig:segan-results} and Fig.~\ref{fig:sewavenet-results}.   Our models trained only on the Valentini dataset are labeled as \textit{SalientL} for the large model and \textit{SalientS} for the small model. The provided data fell into distinct input SNR ranges, so we analyzed them accordingly.

The results show that the clones-extracted salient features outperform SEWaveNet at each input SNR range.  The results match or exceed the performance for SEGAN at each input SNR range as well. 

The speech generated by our system does not attempt to minimize the error with the ground truth and as such, creates significantly different waveforms, shown in Fig.~\ref{fig:waveform-comparison}. As the noise level increases, our system produces phoneme errors and speech-like perturbations from the ground truth instead of `traditional' artifacts that are outside of the speech manifold. The results indicate that these were preferable to the listeners to the `traditional'-sounding artifacts produced by SEWaveNet or SEGAN, as well as the noisy signal itself.

\section{CONCLUSION}
\label{sec:conclusion}

We proposed and evaluated a generative enhancement system based on cloned networks that matched or exceeded the performance of more restrictive generative systems. The generated output has a particular type of degradation that avoids unnatural-sounding artifacts by being trained to produce only speech-like sounds.

There are many possibilities for future work.  As decreasing SNR results in phoneme errors in generative systems, it may be desirable to introduce signal augmentation that, at very low SNR, signals to the listener that the input quality is poor. One approach would be to introduce synthesized background noise with an agreeable and perhaps accurate character.  Along with exploring other architectures, we can also consider the unsupervised case without a decoder.

\bibliographystyle{IEEEtran}
\bibliography{refs19}
%
%
%
%
%
%
%
%
%

\end{sloppy}
\end{document}